\documentclass{article}

\usepackage{graphicx}
\usepackage{amssymb}
\usepackage{cite}

\makeatletter

\def\eqnumsection{
    \@addtoreset{equation}{section}
    \def\theequation{\arabic{section}.\arabic{equation}}
}

\makeatother

\def\D{\mathrm{d}} 
\def\E{\mathrm{e}}
\def\I{\mathrm{i}}

\title{
\begin{flushright}
{\normalsize Yaroslavl State University\\
             Preprint YARU-HE-06/01\\ 
             hep-ph/0605114} \\[20mm]
\end{flushright}
NEUTRINO PROPAGATION IN MAGNETIZED PLASMA
}

\author{
A.V. Kuznetsov and N.V. Mikheev\\[3mm]
{\it Division of Theoretical Physics,} \\
{\it Yaroslavl State (P.G.~Demidov) University,} \\
{\it Sovietskaya 14, 150000 Yaroslavl, Russian Federation}\\[3mm]
{\tt E-mail: avkuzn@uniyar.ac.ru, mikheev@uniyar.ac.ru}
}

\date{}

\begin{document}

\maketitle

\begin{abstract}
The process of neutrino propagation through an active medium 
consisting of magnetic field and plasma is analysed. 
We consider in detail the contribution of a magnetic field $B$
into the neutrino self-energy operator $\Sigma (p)$. The results 
for this contribution were contradictory in the previous literature.  
For the conditions of the early
universe where the background medium consists of a charge-symmetric
plasma, the pure $B$-field contribution to the neutrino dispersion
relation is proportional to $(e B)^2$ and thus comparable to the
contribution of the magnetized plasma.
The neutrino self-energy operator $\Sigma (p)$ is calculated 
also for the case of high-energy neutrinos, which corresponds 
to the crossed field approximation. The probability of the neutrino decay 
$\nu \to e^- W^+$ is calculated from the imaginary part of the 
$\Sigma (p)$ operator. A simple analytical result is obtained for the most 
interesting region of parameters which was not considered earlier.
The external magnetic field contribution into the neutrino magnetic moment 
is calculated. The result obtained corrects the formulas existed earlier. 
We show qualitatively that the effect of ``neutrino spin light'' 
discussed in the literature has no physical region of 
realization because of the medium influence on photon dispersion. 
\end{abstract}
 
\vfill

\begin{center}
{\it Lecture presented at 
the XL PNPI Winter School on Nuclear and Particle Physics} \\ 
 {\it and XII St. Petersburg School on Theoretical Physics,} \\
 {\it St.-Petersburg, Repino, Russia, February 20-26, 2006}
\end{center}

\newpage

\eqnumsection

\section{Introduction}

The presence of matter or electromagnetic fields modifies the
dispersion relation of neutrinos in rather subtle ways because these
elusive particles interact only by the weak force. However,
it was recognized that the feeble matter effect
is enough to affect neutrino flavor oscillations in dramatic ways
because the neutrino mass differences are very
small~\cite{Wolfenstein:1978,Mikheyev:1985}, with practical applications
in physics and astrophysics whenever neutrino oscillations
are important~\cite{Mohapatra:1998rq}.

The presence of external fields will lead to additional modifications
of the neutrino dispersion relation. There is a natural scale for the
field strength that is required to have a significant impact on
quantum processes, i.e.\ the critical value
\begin{equation}
B_e = m_e^2/e \approx 4.41 \times 10^{13}~\textrm{G}\,.
\end{equation}
Note that we use natural units where $\hbar=c=1$ and the
Lorentz-Heaviside convention where
$\alpha=e^2/4\pi\approx1/137$ so that $e\approx0.30>0$ is the
elementary charge, taken to be positive.

There are reasons to expect that fields of such or even larger
magnitudes can arise in cataclysmic astrophysical events such as
supernova explosions or coalescing neutron stars, situations where a
gigantic neutrino outflow should also be expected.  There are two
classes of stars, i.e.~soft gamma-ray repeaters
(SGR)~\cite{Kouveliotou:1999,Hurley:1999} and anomalous x-ray pulsars
(AXP)~\cite{Li:1999,Mereghetti:2002} that are believed to be remnants
of such cataclysms and to be magnetars~\cite{Duncan:1992}, neutron
stars with magnetic fields $10^{14}$--$10^{15}$~G.  The
possible existence of even larger fields of order
$10^{16}$--$10^{17}$~G is subject to
debate~\cite{Bisnovatyi-Kogan:1970, Bisnovatyi-Kogan:1989,
Balbus:1998, Akiyama:2003, Ardeljan:2004}.  The early universe between
the QCD phase transition (${}\sim 10^{-5}$~s) and the nucleosynthesis
epoch (${}\sim 10^{-2}$--$10^{+2}$~s) is believed to be yet another
natural environment where strong magnetic fields and large neutrino
densities could exist simultaneously~\cite{Grasso:2001}.

The modification of the neutrino dispersion relation in a magnetized
astrophysical plasma was studied in the previous
literature~\cite{D'Olivo:1989,Semikoz:1994,Elmfors:1996,
Erdas:1998}.  In particular, a charge-symmetric plasma with $m_e \ll T
\ll m_W$ and $B \lesssim T^2$ was considered for the early-universe
epoch between the QCD phase transition and big-bang nucleosynthesis.
Ignoring the neutrino mass, the dispersion relation for the electron
flavor was found to be~\cite{Elmfors:1996,Erdas:1998}
\begin{eqnarray}
\frac{E}{|{\bf p}|}=1&+&
\frac{\sqrt{2}\,G_{\rm F}}{3}
\left[
-\frac{7\pi^2T^4}{15} 
\left(\frac{1}{m_Z^2} + \frac{2}{m_W^2} \right)
+ \frac{T^2eB}{m_W^2}\,\cos\phi \, + \right. 
\nonumber\\
&+& \left. \, \frac{(eB)^2}{2\pi^2m_W^2}\,
\ln\left(\frac{T^2}{m_e^2}\right)\,\sin^2 \phi \right],
\label{eq:E_Raf}
\end{eqnarray}
where $\bf p$ is the neutrino momentum and $\phi$ is the angle between
$\bf B$ and $\bf p$.  The first term proportional to $G_{\rm F}$
in Eq.~(\ref{eq:E_Raf}) is the
dominating pure plasma contribution~\cite{Notzold:1988}, whereas the
second term is caused by the common influence of the plasma and
magnetic field~\cite{Elmfors:1996}.  The third term is of the second order
in $(eB/T^2)\ll 1$ but was included because of the large logarithmic
factor $\ln(T/m_e)\gg 1$ \cite{Erdas:1998}. The dispersion relation
of Eq.~({\ref{eq:E_Raf}}) applies to both $\nu_e$ and $\bar\nu_e$
without sign change in any of the terms.

The $B$-field induced pure vacuum modification of the neutrino
dispersion relation 
was assumed to be negligible in these papers.

However, recently this contribution was calculated for the same
conditions~\cite{Elizalde:2002,Elizalde:2004}, with an absolutely different result:
\begin{equation}
\frac{\Delta E}{|{\bf p}|}= 
\sqrt{2}\,G_{\rm F}\,
\frac{eB}{8\pi^2}\,\sin^2 \phi \; \E^{-p_\bot^2/(2eB)}\,,
\label{eq:DE_Elizalde}
\end{equation}
where $p_\bot$ is the momentum component perpendicular to the
$B$-field. It is easy to check that this would be the dominant
$B$-field induced contribution by far and thus would lead to important
consequences for neutrino physics in 
media~\cite{Ferrer:2003,Ferrer:2004}.

Because of importance of the question whether the $B$-field contribution 
into the neutrino dispersion relation was dominating or negligible, an 
independent calculation of it was strongly urged.

One more promising effect based on using the neutrino dispersion properties 
in external active medium, the so-called ``neutrino spin light'', 
was proposed in the series of papers~\cite{Lobanov:2003, Studenikin:2005, 
Grigoriev:2005a, Grigoriev:2005b, Lobanov:2005}, however, the medium influence on the 
photon dispersion was not considered there. 

The paper is organized as follows. 
We begin in Secs.~\ref{sec:definition} and~\ref{sec:propagators} with
the technique to calculate the neutrino self-energy operator by using
the charged-lepton, $W$- and $\Phi$-boson propagators in a magnetic field.  In
Sec.~\ref{sec:calculation} we derive explicit results for the neutrino
self-energy operator in the limiting cases of a ``weak field'' $e B
\ll m_\ell^2$ and a ``moderate field'' $m_\ell^2 \ll e B \ll m_W^2$. 
In Sec.~\ref{sec:energy} we find the pure-field 
correction to the neutrino energy and in Sec.~\ref{sec:resonance} 
we study its possible contribution into the resonance condition for 
neutrino oscillations in the supernova interior. 
The probability of the neutrino decay $\nu \to e^- W^+$ 
and the neutrino magnetic moment in an external electromagnetic field
are calculated in Secs.~\ref{sec:decay} and~\ref{sec:moment}.  
In Sec.~\ref{sec:spinlight} we study a question whether 
the effect of ``neutrino spin light'' has a physical region of realization 
with the photon dispersion in medium taken into account,
before concluding in Sec.~\ref{sec:conclusions}.

\section{Definition of the neutrino self-energy operator $\Sigma(p)$} 
\label{sec:definition}

A literature search reveals that calculations of the neutrino
dispersion relation in external $B$-fields have a long
history~\cite{McKeon:1981,Borisov:1985,Erdas:1990}. To compare the
different results we introduce the neutrino self-energy operator
$\Sigma (p)$ that is defined in terms of the invariant amplitude for
the neutrino forward scattering on vacuum fluctuations, $\nu \to \nu$, 
by the relation
\begin{equation}
{\cal M}(\nu\to\nu)=-\bar\nu(p)\,\Sigma (p)\,\nu(p)\,,
\label{eq:sigma_def}
\end{equation}
where $p$ is the neutrino four-momentum. Note that we use the
signature $(+,-,-,-)$ for the four-metric. 

Perturbatively, the matrix element of Eq.~(\ref{eq:sigma_def})
corresponds in the Feynman gauge to the Feynman diagrams shown in Fig. 1 
where double lines denote exact propagators in the external $B$ field.
Put another way, the terms in the self-energy operator correspond to these Feynman
graphs with the external neutrino lines truncated.


\begin{figure}[htb]
\centering
\includegraphics[width=0.65\textwidth]{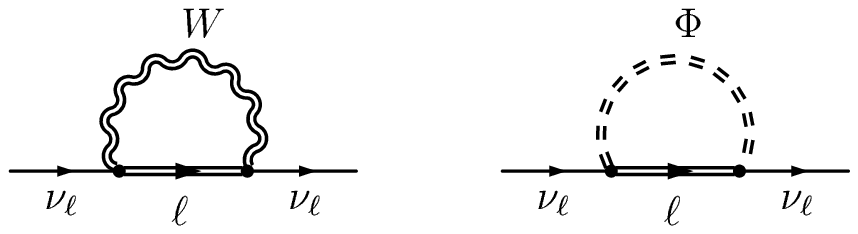}
\begin{quote}
\begin{center}
{\small {\bf Fig. 1.} 
The Feynman diagrams for the $\nu \to \nu$ transition in the Feynman gauge. 
The double lines denote exact
propagators of the charged lepton, the $W$-boson, and the unphysical 
charged scalar $\Phi$-boson in an external magnetic field.
}
\end{center}
\end{quote}
\end{figure}


At first glance, the contribution of the diagram with the scalar field should be 
negligible because of the suppression by the factor $(m_{\ell}/m_W)^2$ arising 
from the coupling of the lepton with the scalar $\Phi$-boson. 
However, as will be shown later, it is essential in some cases. 

\section{Calculation techniques} 
\label{sec:propagators}

The calculation techniques for quantum processes in external electromagnetic fields
based on exact propagators in the field started from the classical paper by 
J. Schwinger~\cite{Schwinger:1951} and was developed by A.~Nikishov, V. Ritus, A. Shabad, 
V. Skobelev et al. For a recent review see e.g.~\cite{Kuznetsov:2003}. 

The exact propagator for the charged lepton $\ell$ (for definiteness we 
take $Q_\ell = - e < 0$)
in a constant and uniform magnetic field can be expressed as
\begin{equation}
S^{F}(x,y)  = \E^{\mbox{\normalsize $\I \Omega (x,y)$}}\,S (x-y)\,,
\label{eq:S_L}
\end{equation}
and similarly for the $W$- and $\Phi$-bosons:
\begin{eqnarray}
G^F_{\rho \sigma} (x,y) &=& \E^{\mbox{\normalsize $\I \Omega (x,y)$}} \,
G_{\rho \sigma} (x-y)\,,
\label{eq:S_W}\\
D^F (x,y) &=& \E^{\mbox{\normalsize $\I \Omega (x,y)$}} \,
D (x-y)\,.
\label{eq:S_phi}
\end{eqnarray}
where $S (x-y)$, $G_{\rho \sigma} (x-y)$, $D (x-y)$ are the translationally 
and gauge invariant parts of the propagators.  

The phases $\Omega (x,y)$ being identical for all propagators are translationally 
and gauge non-invariant, but they cancel in the two-vertex loop: 
\begin{equation}
\Omega (x,y)  + \Omega (y,x) = 0\,.
\label{eq:phase=0}
\end{equation}
The Fourier transforms of the translationally invariant parts of the propagators 
are defined by:
\begin{eqnarray}
S (X) &=& \int \frac{\D^4 q}{(2 \pi)^4} \, S (q) \, \E^{- \I q X}\,,
\label{eq:S(X)}\\
G_{\rho \sigma} (X) &=& \int \frac{\D^4 q}{(2 \pi)^4} \, G_{\rho\sigma} (q)\,\E^{- \I q X}\,,
\label{eq:G(X)}\\
D (X) &=& \int \frac{\D^4 q}{(2 \pi)^4} \, D (q) \, \E^{- \I q X}\,.
\label{eq:D(X)}
\end{eqnarray}

For the Fourier transform $S (q)$ of the translationally invariant
part of the lepton propagator one obtains in the
Fock proper-time formalism:
\begin{eqnarray}
S (q) &=& \int_0^{\infty}\!\! 
\frac{\D s}{\cos (\beta s)}\,
\exp\left[- \I s \left(m_\ell^2- q_{\|}^2 
+ \frac{\tan (\beta s)}{\beta s}\,q_{\perp}^2 \right) \right] 
\nonumber\\
&\times& \biggl\{\left[(q \gamma)_{\|} + m_\ell\right] \left[
\cos (\beta s) - \frac{ (\gamma \varphi \gamma)}{2} \,
\sin (\beta s) \right]
-\frac{(q \gamma)_{\perp}}{\cos (\beta s)}\biggr\},
\label{eq:S(q)}
\end{eqnarray}
where $\beta = e B$ and $m_\ell$ is the lepton mass, 
$\varphi_{\alpha\beta} = F_{\alpha\beta}/B$ is the 
dimensionless field tensor. 
The Lorentz indices of four-vectors and tensors
within parentheses are contracted consecutively, e.g. 
$(\gamma \varphi \gamma) = \gamma^{\alpha} \varphi_{\alpha\beta} \gamma^{\beta}$.
In the frame where the $B$ field is directed along the 3d axis,
four-vectors with the indices $\bot$ and $\|$ belong to the Euclidean
$\{1, 2\}$-subspace and the Minkowski $\{0, 3\}$-subspace,
correspondingly. For example, $p_\bot=(0,p_1,p_2,0)$ and
$p_\|=(p_0,0,0,p_3)$.  For any four-vectors $X$ and $Y$ we write, 
both in the invariant form and in the above-mentioned frame:
\begin{eqnarray}
(XY)_{\|} &=& (X \,\tilde \varphi \tilde \varphi \,Y) = 
X_0 Y_0 - X_3 Y_3\,, 
\nonumber\\
(XY)_{\perp} &=& (X \,\varphi \varphi \,Y) =  X_1 Y_1 + X_2 Y_2\,, 
\nonumber\\
(XY) &=& (XY)_{\|} - (XY)_{\perp}\,.
\label{eq:XY}
\end{eqnarray}

Similarly, for the $W$- and $\Phi$-boson propagators 
in the Feynman gauge we have:
\begin{eqnarray}
G_{\rho \sigma} (q) &=& - \int_0^{\infty} 
\frac{\D s}{\cos (\beta s)} 
\exp\left[- \I s \left(m_W^2- q_{\|}^2 
+ \frac{\tan (\beta s)}{\beta s}\,q_{\perp}^2 \right) \right] 
\nonumber\\
&\times& \biggl[
(\tilde\varphi \tilde\varphi)_{\rho \sigma} 
- (\varphi \varphi)_{\rho \sigma} \, \cos (2 \beta s)
- \varphi_{\rho \sigma} \, \sin (2 \beta s) \biggr] ,
\label{eq:G(q)}\\
D (q) &=& \int_0^{\infty} 
\frac{\D s}{\cos (\beta s)} 
\exp\left[- \I s \left(m_W^2- q_{\|}^2 
+ \frac{\tan (\beta s)}{\beta s}\,q_{\perp}^2 \right) \right] .
\label{eq:D(q)}
\end{eqnarray}

Magnetic fields existing in
Nature probably are always weak compared with the critical field for
the $W$-boson, $B_W = m_W^2/e \simeq 10^{24}\,\textrm{G}$.  Therefore, the
$W$-propagator can be expanded in powers of $\beta$ as a small
parameter:
\begin{eqnarray}
G_{\rho \sigma} (q) &=&
- \I \, \frac{g_{\rho \sigma}}{q^2 - m_W^2} \, - 
\, \beta \, \frac{2 \, \varphi_{\rho \sigma}}{(q^2 - m_W^2)^2} \, 
\label{eq:G(q)weak}\\
&+& 
\I\, \beta^2 \biggl[g_{\rho \sigma} \left(\frac{1}{(q^2 - m_W^2)^3} + 
\frac{2 \, q_{\perp}^2}{(q^2 - m_W^2)^4} \right) 
+ 
4 \, (\varphi \varphi)_{\rho \sigma} \,\frac{1}{(q^2 - m_W^2)^3} 
\biggr]
+ {\cal O}(\beta^3)\,.
\nonumber
\end{eqnarray}

Likewise, the asymptotic expression for the lepton propagator $S (q)$
is realised when the field strength is the smallest dimensional parameter, 
$\beta \ll m_\ell^2 \ll m_W^2$.  In this ``weak field approximation'' the
charged-lepton propagator can be expanded as~\cite{Chyi:2000}:
\begin{eqnarray}
S (q) &=&
\I \, \frac{(q \gamma) + m_\ell}{q^2 - m_\ell^2} \, + 
\, \beta \, \frac{(q \gamma)_{\|} + m_\ell}{2 (q^2 - m_\ell^2)^2} 
\,(\gamma \varphi \gamma)
\nonumber\\
&+&\beta^2 \, \frac{2 \, \I
\left[(q_{\|}^2 - m_\ell^2) \, (q \gamma)_{\perp} 
- q_{\perp}^2 \, ((q \gamma)_{\|} + m_\ell) \right]}
{(q^2 - m_\ell^2)^4} 
+ {\cal O}(\beta^3)\,.
\label{eq:S(q)weak}
\end{eqnarray}

One can see that the contribution of the region of
small virtual momenta $q^2 \sim m_\ell^2 \ll m_W^2$ is enhanced in
each succeeding term. If the propagator is used for a ``moderate
field'', $m_\ell^2 \ll \beta \ll m_W^2$, the expansion is not
applicable and the exact propagator~(\ref{eq:S(q)}) must be taken.

\section{Calculation of the $\Sigma(p)$ operator}
\label{sec:calculation}

From the $\cal S$ matrix element for the transition $\nu \to \nu$
corresponding to the Feynman diagrams of Fig. 1,  
one extracts the amplitude $\cal M$. 
The neutrino self-energy operator can be written in the form:
\begin{eqnarray}
\Sigma (p) &=& - \frac{\I\,g^2}{2} \left[ \gamma^\alpha \, L \, 
J^{(W)}_{\alpha \beta} (p) \, \gamma^\beta \, L \right .
\nonumber\\
&+& \left . \frac{1}{m_W^2}
\left(m_\ell R - m_\nu L \right) J^{(\Phi)} (p) 
\left(m_\ell L - m_\nu R \right) \right] ,
\label{eq:Sigma_J}
\end{eqnarray}
where $g$ is the electroweak $SU (2)$ coupling constant of the
standard model, $L=\frac{1}{2}(1-\gamma_5)$ and $R=\frac{1}{2}(1+\gamma_5)$ 
are the left-handed and right-handed projection operators, and  
\begin{equation}
J^{(W)}_{\alpha \beta} (p) = \int \, \frac{\D^4 q}{(2 \pi)^4} \,S (q) \, 
G_{\beta \alpha} (q-p) \,, \quad
J^{(\Phi)} (p) = \int \, \frac{\D^4 q}{(2 \pi)^4} \,S (q) \, D (q-p) \,.
\label{eq:J's}
\end{equation}

It is convenient to express the structure of the $\Sigma(p)$ operator 
in an external magnetic field in terms of the coefficients 
${\cal A}_L$, ${\cal B}_L$, ${\cal C}_L$, ${\cal A}_R$, etc.

\begin{eqnarray}
\Sigma(p)&=&
\left[{\cal A}_L\,(p\gamma) + {\cal B}_L \,e^2 \left(p \tilde F \tilde F \gamma \right) + 
{\cal C}_L \,e \left(p \tilde F \gamma \right) \right]\, L 
\nonumber\\[2mm]
&+& \left[{\cal A}_R\,(p\gamma) + {\cal B}_R\,e^2 \left(p \tilde F \tilde F \gamma \right) + 
{\cal C}_R \, e \left(p \tilde F \gamma \right) \right]\, R 
\nonumber\\[2mm]
&+& 
m_\nu \, \left[{\cal K}_1 + \I \, {\cal K}_2 \, e \left(\gamma F \gamma \right) \right]
\,, 
\label{eq:sigma_cros}
\end{eqnarray}
where $F$ is the external field tensor, and $\tilde F$ is its dual. 

Here, the coefficients ${\cal A}_R$, ${\cal B}_R$, ${\cal C}_R$, 
${\cal K}_{1,2}$ are originated from the Feynman diagram of Fig. 1 
with the scalar $\Phi$-boson, 
while the coefficients ${\cal A}_L$, ${\cal B}_L$, ${\cal C}_L$ 
contain the contributions of both diagrams. 

The coefficients ${\cal A}_L$, ${\cal A}_R$ and ${\cal K}_1$, 
being ultraviolet divergent, do not have independent meanings, 
because they do not give contributions into the real neutrino energy in external field 
at the one-loop level. 
They are absorbed by the neutrino wave-function and mass renormalization.
The coefficients ${\cal B}_R$, ${\cal C}_R$ are suppressed by the factor $(m_{\nu}/m_W)^2$. 
The coefficient ${\cal K}_2$ is suppressed by the factor $(m_{\ell}/m_W)^2$, 
see, however, Sec.~\ref{sec:moment}. 
Thus, the coefficients ${\cal B}_L$, ${\cal C}_L$ are of the most interest. 

We present our results for the ${\cal B}_L$ and ${\cal C}_L$ coefficients 
of the $\Sigma(p)$ operator~(\ref{eq:sigma_cros}) in Table 1 where the results of previous 
authors~\cite{McKeon:1981,Borisov:1985,Erdas:1990} and of the 
above-mentioned papers~\cite{Elizalde:2002,Elizalde:2004} are also shown.

\section{Neutrino energy in a magnetic field}
\label{sec:energy}

Solving the equation for the neutrino dispersion in a magnetic field
(for~$m_\nu = 0$)

\begin{equation}
\textrm{det} \, \left|(p \gamma) - {\cal B}_L \, e^2 (p \tilde F \tilde F \gamma) \,L
- {\cal C}_L \, e (p \tilde F \gamma) \,L \right| = 0 
\,,
\label{eq:det}
\end{equation}
where the leading terms with ${\cal B}_L$, ${\cal C}_L$
are only included, one obtains for the neutrino energy in the field:

\begin{equation}
\frac{E}{|{\bf p}|} = 1 +
\left({\cal B}_L + \frac{{\cal C}_L^2}{2}\right) (e B)^2 \, \sin^2 \phi \,.
\label{eq:E/p_CB}
\end{equation}
%


\begin{quote}
\begin{center}
{\small {\bf Table 1.} Coefficients in Eq.~(\ref{eq:sigma_cros})
for the neutrino self-energy operator $\Sigma(p)$ in an external
$B$-field.}
\end{center}
\end{quote}
\vspace{3mm}
{\normalsize 
\begin{table*}
\begin{tabular}{lccc}
\hline\\
Authors&Field strength
&$\displaystyle{{\cal B}_L\times\frac{\sqrt{2}\,\pi^2}{G_{\rm F}}}$
&$\displaystyle{{\cal C}_L\times\frac{\sqrt{2}\,\pi^2}{G_{\rm F}}}$
\\[2ex]
\hline\\
McKeon (1981)&Moderate&0&
$+3$\\[2ex]
Borisov et al.\ (1985)&Arbitrary&
---&$\displaystyle{+\frac{3}{4}}$
\\[2ex]
Erdas, Feldman (1990)&Moderate&
$\displaystyle{-\frac{1}{3m_W^2}\,
\left(\ln\frac{m_W^2}{m_\ell^2}+\frac{3}{4}\right)}$&0
\\[4ex]
Elizalde et al.\ (2002)&Moderate&
$\displaystyle{+\frac{1}{2 eB}}$&
$\displaystyle{-\frac{1}{2}}$
\\[4ex]
Elizalde et al.\ (2004)
&Moderate&
$\displaystyle{+\frac{1}{4 eB}\,
\E^{-p_\bot^2/(2eB)}}$&
$\displaystyle{-\frac{1}{4}\,
\E^{-p_\bot^2/(2eB)}}$
\\[4ex]
Our result (2005)&Weak&
$\displaystyle{-\frac{1}{3m_W^2}\,
\left(\ln\frac{m_W^2}{m_\ell^2}+\frac{3}{4}\right)}$&
$\displaystyle{+\frac{3}{4}}$
\\[4ex]
Our result (2005)&Moderate&
$\displaystyle{-\frac{1}{3m_W^2}\,
\left(\ln\frac{m_W^2}{eB}+2.54\right)}$&
$\displaystyle{+\frac{3}{4}}$
\\[2ex]
\hline
\end{tabular}
\end{table*}

It can be seen that the ${\cal B}_L$ coefficient gives the main contribution into 
the neutrino energy, because the value ${\cal C}_L^2/{\cal B}_L \sim G_{\rm F} m_W^2$ 
appears to be of the 
order of the fine-structure constant $\alpha \simeq 1/137$, thus leading us beyond the frame of the 
one-loop approximation.

Our results strongly disagree with those by E.~Elizalde e.a.~\cite{Elizalde:2002,Elizalde:2004}.
We think that the
disagreement arises because these authors use only one lowest Landau
level in the charged-lepton propagator in the case of moderate field
strengths which they call ``strong fields.''  However, the
contributions of the next Landau levels appear to be of the same order as
the ground-level contribution~\cite{Kuznetsov:2006} because in the integration over the
virtual lepton four-momentum in the loop the region $q^2 \sim
m_W^2 \gg \beta$ appears to be essential.  

We confirm the assumption of Refs.~\cite{D'Olivo:1989,Elmfors:1996},
that the pure magnetic field contribution into the neutrino energy does not exceed the 
plasma contribution.

For relatively weak field $e B \ll m_e^2$ we find the pure-field correction to 
the electron neutrino energy in a magnetic field and plasma, rewriting Eq.~(\ref{eq:E_Raf}) 
as follows: 
\begin{eqnarray}
\frac{E}{|{\bf p}|}=1 &+&
\frac{\sqrt{2}\,G_{\rm F}}{3}\,
\left[-\frac{7\pi^2T^4}{15}
\left(\frac{1}{m_Z^2} + \frac{2}{m_W^2} \right)
+\frac{T^2eB}{m_W^2}\,\cos\phi \right. 
\nonumber\\[2mm]
&+& \left. \frac{(e B)^2}{2\pi^2m_W^2}
\,\sin^2 \phi 
\left( \ln \frac{T^2}{m_e^2} - \ln \frac{m_W^2}{m_e^2} - 
\frac{3}{4}\right)
\right]\,.
\label{eq:E/p_final}
\end{eqnarray}

It is seen from the last terms that 
the pure magnetic field contribution to the neutrino dispersion 
is proportional to $(e B)^2$ and thus comparable to the 
contribution of the magnetized plasma. It is interesting to note 
that the contributions of plasma and of pure magnetic field 
into Eq.~(\ref{eq:E/p_final}), containing the 
electron mass singularities $\sim \ln m_e$, exactly cancel each other. 
It looks suspicious that the singularity remains in the case of 
a moderate field. Possibly it means that the plasma term $\sim (e B)^2$ 
obtained in Ref.\cite{Erdas:1998} is valid in the weak field case only.

\section{Field-induced resonance transition $\nu_{\tau, \mu} \to \nu_e$}
\label{sec:resonance}

There exists a long-standing problem in the supernova explosion modelling, 
of searches of an energy transfer mechanism 
from the neutrino outflow to the stalled 
shock wave for its revival~\cite{Wilson:1985,Bethe:1985}.

If a strong magnetic field is generated inside the exploding 
supernova~\cite{Bisnovatyi-Kogan:1970,Bisnovatyi-Kogan:1989,Balbus:1998,Akiyama:2003,Ardeljan:2004}, 
conditions could arise for the resonance enhancement of the neutrino oscillations
$\nu_{\mu,\tau} \to \nu_e$ (the effect similar to MSW~\cite{Wolfenstein:1978,Mikheyev:1985}), 
with further $\nu_e$-energy transfer to the stellar matter via 
the URCA processes. 

With the magnetic field contribution into the difference of the neutrino self-energies 
$\Delta E = E_{\nu_\ell} - E_{\nu_e} (\ell = \mu, \tau)$,  
when $m_e^2 \ll e B \ll m_\ell^2 \ll m_W^2$,  
the resonance condition for the $\nu_\ell \to \nu_e$ oscillation is
\begin{equation}
\frac{\Delta m_\nu^2}{2 E} \, \cos 2 \theta \,+ \,
\frac{G_F \,(e B)^2}{3 \sqrt{2} \pi^2} \,\frac{E \, \sin^2 \phi}
{m_W^2} \left( \ln \frac{m_\ell^2}{e B} + 1.8 \right) 
- \frac{\sqrt{2} \, G_F \, \rho \, Y_e}{m_N} = 0\,,
\label{eq:res_cond}
\end{equation}
where 
$\theta$ is the mixing angle in the $\nu_\ell, \, \nu_e$ system,
$\rho$ is the matter density, 
$Y_e$ is the electron fraction with respect to nucleons, 
$m_N$ is the nucleon mass. 
It is remarkable that the sign of the field-induced neutrino self-energy 
difference is favorable for the resonance appearance. 

Neglecting the neutrino masses, 
we obtain the following equation for evaluation of the magnetic field strength 
providing the resonance transition $\nu_\tau \to \nu_e$:
\begin{equation}
B_{17}^2 \, (1 - 0.10 \times \ln B_{17}) \simeq 2.5 \times 10^2 \times 
\frac{\rho_7 \, Y_{0.5}}{E_{10}}\,,
\label{eq:res_cond_B}
\end{equation}
where 
$B_{17} = B/(10^{17}\,\textrm{G})$, 
$\rho_7 = \rho/(10^{7}\,\textrm{g/cm}^3)$, 
$Y_{0.5} = Y_e/{0.5}$, 
$E_{10} = E/(10\,\textrm{MeV})$. 

The analysis shows that for realisation of the resonance 
transition $\nu_{\tau, \mu} \to \nu_e$ the field strength is necessary 
$B \gtrsim 10^{18}\,\textrm{G}$, far exceeding the maximal magnetic field 
strength which is believed 
to arise inside the exploding supernova.

\section{Neutrino decay $\nu \to e^- W^+$ in external \\electromagnetic field}
\label{sec:decay}

One more interesting result which can be extracted from the neutrino self-energy operator 
is the probability of the 
neutrino decay $\nu \to e^- W^+$ in an external electromagnetic field~\cite{Borisov:1985}.
It is defined by the imaginary part of the amplitude:
\begin{equation}
w (\nu \to e^- W^+) = \frac{1}{E}\, \textrm{Im} \, {\cal M}(\nu\to\nu)
\,.
\label{eq:w_def}
\end{equation}
In the case of high neutrino energy, which is only interesting for this process, 
the probability is expressed in terms of the dynamical parameter $\chi$
\begin{equation}
\chi^2 = \frac{e^2 \, (p F F p)}{m_W^6}
\,,
\label{eq:chi}
\end{equation}
and the crossed-field approximation is avaliable.
The calculation technique for quantum processes in an external crossed field 
was developed by A.~Nikishov and V. Ritus. 

In the paper~\cite{Borisov:1985} the probability was written in a general form and 
in the two limiting cases: 
$\chi^2 \ll \lambda = m_e^2/m_W^2 \simeq 4 \cdot 10^{-11}$, 
and $\chi \gg 1$. 
However, rewriting the $\chi$ parameter in the form
\begin{equation}
\chi^2 \simeq 3 \cdot 10^{-3}  
\left(\frac{B}{B_e}\right)^2
\left(\frac{E}{10^{20}\,{\rm eV}}\right)^2
\,,
\label{eq:chi_num}
\end{equation}
one can see that for very wide regions of the magnetic field value and the 
neutrino energy the $\chi$ parameter belongs to the interval
\begin{equation}
\lambda \ll \chi^2 \ll 1
\,,
\label{eq:interval}
\end{equation}
which should be much more interesting.

Our result for the decay probability in the case $\lambda \ll \chi^2 \ll 1$ is:
\begin{eqnarray}
w &=& \frac{2\,G_{\rm F}\,m_W^4\,\chi^2}{3\,\sqrt{2}\,\pi\,E} \left[\frac{1 - ({\bf v s})}{2} +
\left(\frac{3}{2}\,\frac{m_\nu}{m_W}\,\chi + \frac{m_\nu}{2E \tan \phi} \right) ({\bf t s})
\right.
\nonumber\\[2mm]
&+& \left. \frac{m_\nu^2}{2 m_W^2}\,\frac{1 + ({\bf v s})}{2} 
\right]
\,,
\label{eq:w_res}
\end{eqnarray}
where ${\bf v} = {\bf p}/E$ is the neutrino velocity, 
${\bf s}$ is the unit vector of the neutrino spin direction, 
and ${\bf t} = ({\bf n} \times ({\bf B} \times {\bf n}))/(B \sin \phi)$ 
is the unit vector lying in the plane of ${\bf B}$ and 
${\bf n} = {\bf p}/|{\bf p}|$.
The term with ${\bf t}$ exists in the case if the neutrino has a transversal 
polarization. 

The last term in~(\ref{eq:w_res}) is provided by the contribution 
of the Feynman diagram in Fig. 1 with the charged scalar 
$\Phi$-boson. It should be noted that the probability is not positively defined without it, when  
the angle between ${\bf s}$ and ${\bf n}$ is small, but not equal to zero exactly.

\section{Field contribution into the neutrino magnetic moment}
\label{sec:moment}

One more case when the contribution of the diagram with the charged scalar 
$\Phi$-boson is essential, is the field contribution into the magnetic moment 
$\mu_{\nu_\ell}$ of the neutrino $\nu_\ell$. The neutrino magnetic moment 
is expressed in the coefficients of the self-energy operator~(\ref{eq:sigma_cros}) 
as follows:
\begin{eqnarray}
\mu_{\nu_\ell} = \frac{e \,m_\nu}{2} \left({\cal C}_L - {\cal C}_R + 4 \, {\cal K}_2\right) \,.
\label{eq:magmom_def}
\end{eqnarray}

In the limiting case $\chi^2 \ll \lambda = m_\ell^2/m_W^2$ we obtain:
\begin{eqnarray}
\mu_{\nu_\ell} \simeq \mu_{\nu_\ell}^{(0)} \left[ 1 + \frac{4}{3} \, \chi^2 
\left(\ln \frac{1}{\lambda} - 3 \, + \, \frac{1}{3} \right) \right] \,, 
\label{eq:munu}
\end{eqnarray}
where $\mu_{\nu_\ell}^{(0)}$ is the neutrino magnetic moment 
in vacuum~\cite{Lee:1977,Fujikawa:1980}
\begin{equation}
\mu_{\nu_\ell}^{(0)} = \frac{3e\,G_{\rm F} m_\nu}{8\pi^2\sqrt{2}}\,.
\label{eq:munu0}
\end{equation}

The leading term of the field correction $\sim \chi^2$ in~(\ref{eq:munu}) with 
a big logarythm coinsides with the result of Ref.~\cite{Borisov:1985} where the 
post log terms were not taken into account. 
The last term $(1/3)$ in Eq.~(\ref{eq:munu}) provided by the $\Phi$-boson, is relatively small, 
but it is not parametrically suppressed, as it was treated earlier.

\section{No ``neutrino spin light'' because of photon dispersion in medium}
\label{sec:spinlight}

The influence of an active medium on the neutrino dispersion was exploited 
in the recent series of papers by A.~Studenikin et 
al.~\cite{Lobanov:2003, Studenikin:2005, Grigoriev:2005a, Grigoriev:2005b, Lobanov:2005}, 
where the so-called effect of ``neutrino spin light'' was discovered. 

The idea was based on the additional Wolfenstein energy~\cite{Wolfenstein:1978} 
acquired by a left-handed neutrino in medium:
\begin{eqnarray}
E_{\nu_L} &\simeq& E_0 + \frac{G_{\rm F} \,N}{\sqrt{2}}
\left(1 + 4 \, \sin^2 \theta_{\rm W}\right) \,, 
\label{eq:EnuL}\\
E_{\nu_R} &\simeq& E_0 \,,
\label{eq:EnuR}
\end{eqnarray}
where $N$ is the number density of background electrons. 

Given the effective $\nu_L \nu_R \gamma$ vertex caused by the 
neutrino magnetic moment, the decay became possible, in those 
authors' opinion: 
\begin{equation}
\nu_L \to \nu_R + \gamma \,,
\label{eq:nuLRgamma}
\end{equation}
with the photon emission which was called ``the neutrino spin light''.

As is seen from the papers~\cite{Lobanov:2003, Studenikin:2005, Grigoriev:2005a, 
Grigoriev:2005b, Lobanov:2005}, the authors made 
the kinematical analysis and calculations of the decay probability and other 
observables, with taking account of the neutrino dispersion in medium, but 
considering the photon created as it was sterile with respect to the medium 
influence, and had the vacuum dispersion, $\omega = |{\bf k}|$. 
However, it is well-known, that medium modifies essentially the photon dispersion 
to the form $\omega = |{\bf k}|/n$, where $n \ne 1$ is the refractive index. 
Having in mind possible astrophysical applications, it is worthwhile to consider 
the astrophysical plasma as a medium, where the photon acquires properties 
of a plasmon~\cite{Weldon:1982, Braaten:1991, Altherr:1994}. 
The dispersion curves for the transversal and longitudinal plasmon are 
depicted in Fig. 2. 


\begin{figure}[htb]
\centering
\includegraphics[width=0.6\textwidth]{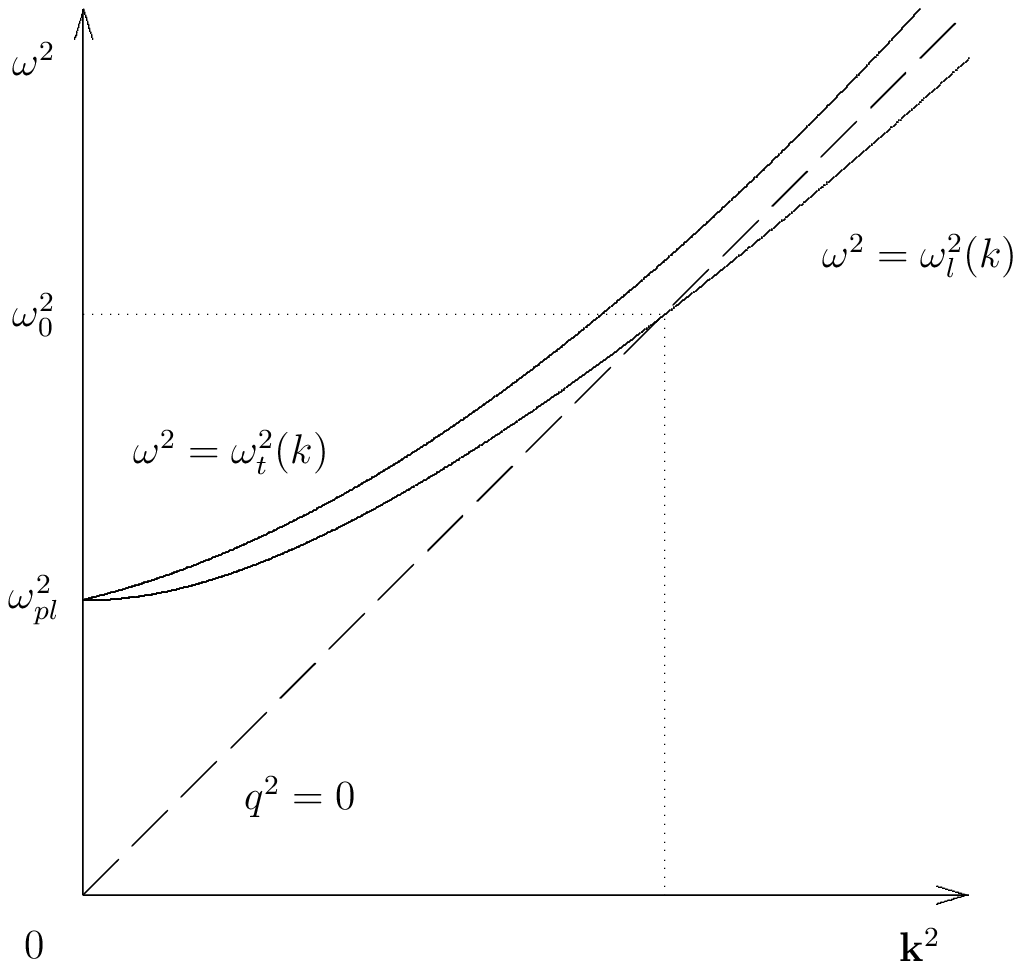}
\begin{quote}
\begin{center}
{\small {\bf Fig. 2.} Dispersion curves 
for the transversal plasmon $\omega^2=\omega^2_t(k)$ (upper solid line),
for the longitudinal plasmon $\omega^2=\omega^2_l(k)$ (lower solid line), 
and for the vacuum photon $q^2 = 0$ (dashed line).
}
\end{center}
\end{quote}
\end{figure}


The deviation of the plasmon dispersion in dense matter from the vacuum one is defined 
by the so-called plasmon frequency which is for the relativistic case:
\begin{equation}
\omega_{\rm pl} = \left(\frac{4 \, \alpha}{3\, \pi} \right)^{1/2} 
\left(3\, \pi^2 \, N \right)^{1/3} \simeq 
0.73 \times 10^{7} \,{\rm eV}
\left(\frac{N}{10^{37} \, {\rm cm}^{-3}}\right)^{1/3}.
\label{eq:omega_pl}
\end{equation}
It should be compared with the Wolfenstein energy defining the neutrino 
dispersion in medium
\begin{equation}
\Delta E_{\rm W} = \frac{G_{\rm F} \,N}{\sqrt{2}}
\left(1 + 4 \, \sin^2 \theta_{\rm W}\right) \simeq 
1.2 \; {\rm eV}
\left(\frac{N}{10^{37} \, {\rm cm}^{-3}}\right).
\label{eq:DeltaE_W}
\end{equation}
Here, the scale of the electron number density is taken, which is typical for the 
interior of a neutron star. For smaller densities, the value $\omega_{\rm pl}$ 
exceeds $\Delta E_{\rm W}$ even much greater. 

As can be seen from the dispersion plot, the 4-momentum of the transversal plasmon is 
always timelike, $\omega^2 > {\bf k}^2 \, (n < 1)$. 
It means that this plasmon has an 
effective ``mass'' which is much greater than the energy benefit caused by the 
neutrino dispersion. So, the decay $\nu_L \to \nu_R \gamma_t$ is 
kinematically forbidden. 
This is also true for the decay $\nu_L \to \nu_R \gamma_l$ in the region 
where the 4-momentum of the longitudinal plasmon is timelike. 

On the other hand, in the region where the 4-momentum 
of the longitudinal plasmon is spacelike, $\omega^2 < {\bf k}^2 \, (n > 1)$, 
the decay $\nu_L \to \nu_R \gamma_l$ is kinematically allowed due to the plasmon 
dispersion (the neutrino Cerenkov process), and the contribution of the neutrino 
dispersion into this effect is negligibly small. 
It should be mentioned also, that the longitudinal plasmon being created in that 
region, is unstable because of the Landau damping, and the neutrino energy 
is not transformed into the ``light'' radiation, but in fact into the energy 
of excitation of plasma electrons. 
Thus, the effect of ``neutrino spin light'' has no physical region of realization.

\section{Conclusions}
\label{sec:conclusions}

\begin{itemize}
\item
We have calculated the neutrino self-energy operator $\Sigma (p)$ in the
presence of a magnetic field $B$.  In particular, we have considered the
weak-field limit $e B \ll m_\ell^2$, 
and a ``moderate field'' case, $m_\ell^2 \ll e B \ll m_W^2$.
Our results strongly disagree with those by E.~Elizalde e.a.~\cite{Elizalde:2002,Elizalde:2004}.
We confirm the assumption by J.~C.~D'Olivo e.a.~\cite{D'Olivo:1989} and by 
P.~Elmfors e.a.~\cite{Elmfors:1996}, 
that the pure magnetic field contribution into the neutrino energy does not exceed the 
plasma contribution.
\item
Applying the possible field-induced resonance enhancement of the neutrino oscillations 
to the problem of the supernova shock wave revival, 
we show that the field strength is necessary for this, 
far exceeding the maximal magnetic field strength which is believed 
to arise inside the exploding supernova.
\item
Using the imaginary part of the neutrino self-energy operator, we 
have calculated the probability of the 
neutrino decay $\nu \to e^- W^+$ in an external electromagnetic field. 
We have considered in part the most interesting region of parameters which was not analysed earlier: 
$\lambda \ll \chi^2 \ll 1$. 
We have shown that the contribution of the Feynman diagram with the charged scalar 
$\Phi$-boson is essential, because the probability is not positively defined without it.
\item
We have calculated the external electromagnetic field contribution into the magnetic moment 
$\mu_{\nu_\ell}$ of the neutrino $\nu_\ell$. 
We have shown that the contribution of the $\Phi$-boson diagram is relatively small, 
but it is not parametrically suppressed.
\item
We have shown qualitatively that the effect of ``neutrino spin light''~\cite{Lobanov:2003, 
Studenikin:2005, Grigoriev:2005a, Grigoriev:2005b, Lobanov:2005} has no 
physical region of realization because of the photon dispersion in medium. 
\end{itemize}

\section*{Acknowledgements}

The authors are grateful to G.~G.~Raffelt and L.~A.~Vassilevskaya for collaboration 
and to V.~A.~Rubakov for useful discussion. 
A.~K. is grateful to the Organizers of the XL PNPI Winter School on Nuclear and 
Particle Physics
and of the XII St. Petersburg School on Theoretical Physics for the invitation 
to present this lecture and for warm hospitality. 

The work was supported in part 
by the Russian Foundation for Basic Research under the Grant No. 04-02-16253, 
and by the Council on Grants by the President of Russian Federation 
for the Support of Young Russian Scientists and Leading Scientific Schools of 
Russian Federation under the Grant No. NSh-6376.2006.2.



\end{document}